\documentstyle[12pt]{article}

  \def\pp{{\mathchoice
          {%displaystyle
              \kern 1pt%
              \raise 1pt
              \vbox{\hrule width5pt height0.4pt depth0pt
                    \kern -2pt
                    \hbox{\kern 2.3pt
                          \vrule width0.4pt height6pt depth0pt
                          }
                    \kern -2pt
                    \hrule width5pt height0.4pt depth0pt}%
                    \kern 1pt
           }
            {%textstyle
              \kern 1pt%
              \raise 1pt
              \vbox{\hrule width4.3pt height0.4pt depth0pt
                    \kern -1.8pt
                    \hbox{\kern 1.95pt
                          \vrule width0.4pt height5.4pt depth0pt
                          }
                    \kern -1.8pt
                    \hrule width4.3pt height0.4pt depth0pt}%
                    \kern 1pt
            }
            {%scriptstyle
              \kern 0.5pt%
              \raise 1pt
              \vbox{\hrule width4.0pt height0.3pt depth0pt
                    \kern -1.9pt  %[e]=0.15pt
                    \hbox{\kern 1.85pt
                          \vrule width0.3pt height5.7pt depth0pt
                          }
                    \kern -1.9pt
                    \hrule width4.0pt height0.3pt depth0pt}%
                    \kern 0.5pt
            }
            {%scriptscriptstyle
              \kern 0.5pt%
              \raise 1pt
              \vbox{\hrule width3.6pt height0.3pt depth0pt
                    \kern -1.5pt
                    \hbox{\kern 1.65pt
                          \vrule width0.3pt height4.5pt depth0pt
                          }
                    \kern -1.5pt
                    \hrule width3.6pt height0.3pt depth0pt}%
                    \kern 0.5pt%}
            }
        }}

  \def\mm{{\mathchoice
                       {%displaystyle
                             \kern 1pt
               \raise 1pt    \vbox{\hrule width5pt height0.4pt depth0pt
                                  \kern 2pt
                                  \hrule width5pt height0.4pt depth0pt}
                             \kern 1pt}
                       {%textstyle
                            \kern 1pt
               \raise 1pt \vbox{\hrule width4.3pt height0.4pt depth0pt
                                  \kern 1.8pt
                                  \hrule width4.3pt height0.4pt depth0pt}
                             \kern 1pt}
                       {%scriptstyle
                            \kern 0.5pt
               \raise 1pt
                            \vbox{\hrule width4.0pt height0.3pt depth0pt
                                  \kern 1.9pt
                                  \hrule width4.0pt height0.3pt depth0pt}
                            \kern 1pt}
                       {%scriptscriptstyle
                           \kern 0.5pt
             \raise 1pt  \vbox{\hrule width3.6pt height0.3pt depth0pt
                                  \kern 1.5pt
                                  \hrule width3.6pt height0.3pt depth0pt}
                           \kern 0.5pt}
                       }}

\catcode`@=11
\def\un#1{\relax\ifmmode\@@underline#1\else
        $\@@underline{\hbox{#1}}$\relax\fi}
\catcode`@=12

\let\du=\du                     % dot-under
\def\a{\alpha}
\def\b{\beta}

\def\d{\delta}

\def\f{\phi}
\def\g{\gamma}
\def\h{\eta}

\def\k{\kappa}
\def\l{\lambda}
\def\m{\mu}
\def\n{\nu}
\def\o{\omega}
\def\p{\pi}
\def\q{\theta}

\def\x{\xi}

\def\O{\Omega}

\def\bo{{\raise-.5ex\hbox{\large$\Box$}}}               % D'Alembertian
\def\pa{\partial}                                       % curly d
\def\de{\nabla}                                         % del
\def\TH{{\raise.2ex\hbox{$\displaystyle \bigodot$}\mskip-4.7mu \llap H \;}}
\def\face{{\raise.2ex\hbox{$\displaystyle \bigodot$}\mskip-2.2mu \llap {$\ddot
        \smile$}}}                                      % happy face
\def\sp#1{{}^{#1}}                              % superscript (unaligned)
\def\abs#1{\left| #1\right|}                    % | |
\def\leftrightarrowfill{$\mathsurround=0pt \mathord\leftarrow \mkern-6mu
        \cleaders\hbox{$\mkern-2mu \mathord- \mkern-2mu$}\hfill
        \mkern-6mu \mathord\rightarrow$}
\def\dvec#1{\vbox{\ialign{##\crcr
        \leftrightarrowfill\crcr\noalign{\kern-1pt\nointerlineskip}
        $\hfil\displaystyle{#1}\hfil$\crcr}}}           % <--> accent
\def\frac#1#2{{\textstyle{#1\over\vphantom2\smash{\raise.20ex
        \hbox{$\scriptstyle{#2}$}}}}}                   % fraction
\def\sfrac#1#2{{\vphantom1\smash{\lower.5ex\hbox{\small$#1$}}\over
        \vphantom1\smash{\raise.4ex\hbox{\small$#2$}}}} % alternate fraction
\def\bfrac#1#2{{\vphantom1\smash{\lower.5ex\hbox{$#1$}}\over
        \vphantom1\smash{\raise.3ex\hbox{$#2$}}}}       % "
\def\afrac#1#2{{\vphantom1\smash{\lower.5ex\hbox{$#1$}}\over#2}}    % "
\def\[{\lfloor{\hskip 0.35pt}\!\!\!\lceil}
\def\]{\rfloor{\hskip 0.35pt}\!\!\!\rceil}

\def\du#1#2{_{#1}{}^{#2}}
\def\ud#1#2{^{#1}{}_{#2}}

\def\ha{{\fracmm12}}
\def\tr{{\rm tr}}
\def\Tr{{\rm Tr}}

\def\un{\underline}
\def\fracmm#1#2{{{#1}\over{#2}}}

\def\low#1{{\raise -3pt\hbox{${\hskip 0.75pt}\!_{#1}$}}}

\newskip\humongous \humongous=0pt plus 1000pt minus 1000pt

\newif\ifdtup

\def\plpl{\raise-2pt\hbox{$\raise3pt\hbox{$_+$}\hskip-6.67pt\raise0.0pt}}
\def\mimi{\raise-2pt\hbox{$\raise3pt\hbox{$_-$}\hskip-6.67pt\raise0.0pt}}

\def\dvm{\raisebox{-.145ex}{\rlap{$=$}}}
\def\DM{{\scriptsize{\dvm}}~~}

\def\ref#1{$\sp{#1)}$}

\parskip=\medskipamount                 % space between paragraphs (LaTeX)
\thicklines                         % thick straight lines for pictures (LaTeX)

\begin{document}

% =========================== UH title page ==========================

\thispagestyle{empty}               % no heading or foot on title page (LaTeX)

\def\border{                                            % UH border
        \setlength{\unitlength}{1mm}
        \newcount\xco
        \newcount\yco
        \xco=-24
        \yco=12
        \begin{picture}(140,0)
        \put(-20,11){\tiny Institut f\"ur Theoretische Physik Universit\"at
Hannover~~ Institut f\"ur Theoretische Physik Universit\"at Hannover~~
Institut f\"ur Theoretische Physik Hannover}
        \put(-20,-241.5){\tiny Institut f\"ur Theoretische Physik Universit\"at
Hannover~~ Institut f\"ur Theoretische Physik Universit\"at Hannover~~
Institut f\"ur Theoretische Physik Hannover}
        \end{picture}
        \par\vskip-8mm}

\def\headpic{                                           % UH heading
        \indent
        \setlength{\unitlength}{.8mm}
        \thinlines
        \par
        \begin{picture}(29,16)
        \put(75,16){\line(1,0){4}}
        \put(80,16){\line(1,0){4}}
      \put(85,16){\line(1,0){4}}
        \put(92,16){\line(1,0){4}}

        \put(85,0){\line(1,0){4}}
        \put(89,8){\line(1,0){3}}
        \put(92,0){\line(1,0){4}}

        \put(85,0){\line(0,1){16}}
        \put(96,0){\line(0,1){16}}
        \put(92,16){\line(1,0){4}}

        \put(85,0){\line(1,0){4}}
        \put(89,8){\line(1,0){3}}
        \put(92,0){\line(1,0){4}}

        \put(85,0){\line(0,1){16}}
        \put(96,0){\line(0,1){16}}
        \put(79,0){\line(0,1){16}}
        \put(80,0){\line(0,1){16}}
        \put(89,0){\line(0,1){16}}
        \put(92,0){\line(0,1){16}}
        \put(79,16){\oval(8,32)[bl]}
        \put(80,16){\oval(8,32)[br]}

        \end{picture}
        \par\vskip-6.5mm
        \thicklines}

\border\headpic {\hbox to\hsize{
\vbox{\noindent ITP--UH -- 26/96 \hfill December 1996 \\
hep-th/9612171 }}}

\noindent
\vskip1.3cm
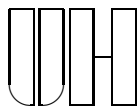
\begin{center}

{\Large\bf Self-Duality and F Theory~\footnote{Talk given at the 4th Nordic 
Meeting on ``Supersymmetric Field and String Theories'' (G\"oteborg, \newline
${~~~~~}$ Sweden, 9--11 September 1996), and at the `mitteldeutsche' Workshop 
(Lutherstadt Wittenberg, \newline ${~~~~~}$ Germany, 29--30 November 1996)}}\\
\vglue.3in

Sergei V. Ketov \footnote{Supported in part by the 
`Deutsche Forschungsgemeinschaft' and the `Volkswagen Stiftung'}

{\it Institut f\"ur Theoretische Physik, Universit\"at Hannover}\\
{\it Appelstra\ss{}e 2, 30167 Hannover, Germany}\\
{\sl ketov@itp.uni-hannover.de}
\end{center}
\vglue.2in
\begin{center}
{\Large\bf Abstract}
\end{center}

The (2,2) world-sheet supersymmetric string theory is discussed from the 
viewpoint of string/membrane unification. The effective field theory in the 
closed string target space is known to be the $2+2$ dimensional (integrable) 
theory of {\it self-dual gravity} (SDG). A world-volume supersymmetrization of 
the Pleba\'nski action for SDG naturally implies the {\it maximal} $N=8$ 
world-volume supersymmetry, while the maximal supersymmetrization of the dual 
covariant K\"ahler-Lorentz-Chern-Simons action for SDG implies gauging a 
self-dual part of the super-Lorentz symmetry in $2+10$ dimensions. The proposed 
$OSp(32|1)$ supersymmetric action for an M-brane may be useful for a fundamental 
formulation of uncompactified F theory, with the self-duality being playing the 
central role both in the world-volume and in the target space of the M-brane.

\newpage

The constructive definition of uncompactified M and F theories is of great 
importance~\cite{schwarz}. By working definition, M theory is a strongly 
coupled type-IIA superstring theory. Stated differently, a weakly coupled 
M theory can be the ten-dimensional type-IIA superstring. It is also known 
that M theory has D-brane (`membrane') non-perturbative (solitonic) degrees of 
freedom, while D=0 branes (`particles`) form a Kaluza-Klein spectrum 
with radius $R\sim g^{2/3}$ where $g$ is the string
coupling~\cite{witten}. It implies that in the infinite string coupling limit, 
$g\to\infty$, a decompactification to eleven spacetime dimensions occurs. The
corresponding low-energy effective action is dictated by supersymmetry, and it 
has to be the eleven-dimensional supergravity which is unique. Similarly, when
starting from the ten-dimensional type-IIB superstring theory, which is self-dual
under the S-duality $SL(2,{\bf Z})$ and is supposed to be a weakly coupled F 
theory, the S-duality relating weak and strong couplings implies twelve  
dimensions in the infinite string coupling limit. The S-duality itself can then 
be naturally explained as the T-duality associated with an extra `torus' 
representing two hidden dimensions~\cite{vafa}. That considerations imply the 
common origin of M-theory and all the ten-dimensional superstrings from a single
F-theory in twelve dimensions.

It was recently conjectured by Banks, Fischler, Shenker and Susskind~\cite{bfss}
that M theory may be the large N limit of a supersymmetric matrix model. They 
used the ten-dimensional supersymmetric U(N) Yang-Mills model dimensionally
reduced to zero dimensions (quantum mechanics~!) and argued that its large N 
limit describes the D=0 branes of M theory. It is known that in the large N limit
 the U(N) gauge group becomes the symmetry of area-preserving diffeomorphisms
to be represented by the loop group $S^1\to SDiff(2)$ on a 2-plane, which can be 
considered as a higher dimensional extension of the W symmetries in 
two-dimensional conformal field theory~\cite{bakas,park}. Since the
area-preserving diffeomorphisms are the natural symmetry of the Hilbert space of 
a membrane (in a light-cone gauge)~\cite{whn}, M and F theories appear to be the
theories of quantized membranes (M-branes). The problem of quantization of
membranes is highly non-trivial, mainly because of their apparent 
non-renormalizability. We should, however, expect the fundamental M-branes to be 
integrable instead, thus giving us a chance for their consistent quantization.
The area-preserving diffeomorphism invariance and the integrability are featured 
by four-dimensional {\it self-dual gravity} (SDG). Therefore, it is conceivable 
that the SDG should play the key role in determining the dynamics of the M-brane 
 \cite{garbsen,jev}. 

More information about the M-brane comes from the observation that the closed 
string theory with $(2,2)$ world-sheet supersymmetry has the interpretation of 
being the theory of SDG~\cite{ov}. The target space of critical $(2,2)$ strings 
is four-dimensional, with the signature $2+2$. It is natural to treat the 
effective field theory of $(2,2)$ strings as an essential part of the 
world-volume theory of the fundamental M-brane. It is known as the 
world-sheet/target space duality principle, suggested by Kutasov and 
Martinec~\cite{kuma} as the basic principle for string/membrane unification. To 
probe M-theory, they applied that principle to the heterotic $(2,1)$ strings 
which, though supersymmetric in the target space, can only be consistently 
defined in lower ($1+2$ or $1+1$) dimensions where the four-dimensional 
self-duality is lost or implicit~\cite{ovh}. Instead, the $(2,2)$ strings are 
intrinsically self-dual. It is the target space supersymmetry that is missing in 
the theory of $(2,2)$ strings. I am going to supersymmetrize the SDG in order to 
get the M-brane action~\cite{garbsen}.

An additional evidence for the possible connection between the $(2,2)$ strings 
and the matrix model of ref.~\cite{bfss} comes from the known reformulation of 
the $(2,2)$ string theory as a quantum topological theory~\cite{berv}. The latter
may apparently be reformulated as the four-dimensional `holomorphic' Yang-Mills 
theory~\footnote{The holomorphic Yang-Mills theory~\cite{korea} is a certain 
topological deformation of the Donaldson-\newline ${~~~~~}$ Witten 
theory~\cite{dwitten} which, in its turn, is a topologically twisted $N=2$ 
supersymmetric four-di-\newline ${~~~~~}$ mensional Yang-Mills theory.} in the 
large N limit~\cite{oog}. 

The effective field theory reproducing closed $(2,2)$ string tree amplitudes is
given by the Pleba\'nski action~\cite{ov},
$$ I_{\rm P}=\int d^{2+2}z\,\left(\fracmm{1}{2}\pa\f\bar{\pa}\f
+\fracmm{\k}{3}\pa\bar{\pa}\f\wedge \pa\bar{\pa}\f\right)~,\eqno(1)$$
where $\k$ is the closed (2,2) string (real) coupling constant, and $\f$ is a 
deformation of the K\"ahler potential $K$ of self-dual (=K\"ahler + Ricci-flat) 
gravity. Given the metric
$$g_{i\bar{j}}=\pa_i\bar{\pa}_{\bar{j}}K=\h_{i\bar{j}}
+4\k \pa_i\bar{\pa}_{\bar{j}}\f~,\quad {\rm where}\quad 
\h_{i\bar{j}}=\left( \begin{array}{cc} 1 & 0 \\ 0 & -1 \end{array}\right)~,
\eqno(2)$$
the equation of motion for the action (1) can be represented in the form
$$ \det g=-1~,\eqno(3)$$
and is invariant under the area-preserving holomorphic diffeomorphisms
({\it cf}. ref.~\cite{pbr}),
$$\pa_i\bar{\pa}_{\bar{j}}K(z,\bar{z})\to \pa_i\x^k(z)
\pa_k\bar{\pa}_{\bar{k}}K(\x,\bar{\x})\bar{\pa}_{\bar{j}}\bar{\x}^{\bar{k}}
(\bar{z})~, \eqno(4)$$
since $\abs{\det(\pa_i\x^k)}=1$ by their definition.

The string world-sheet {\it instanton} contributions are governed by another 
(Maxwell) coupling constant $\l$ which is a phase \cite{klet}. Hence, the 
effective $(2,2)$ closed string coupling constant is {\it complex}, and it seems 
to be rather natural to interpret it as the vacuum expectation value of a complex
 {\it dilaton}, which is usual in string theory. The complex dilaton can be 
described by an anti-self-dual two-form $\o$ satisfying the nilpotency condition
$\o\wedge \o=0$ \cite{ble,ket}. There is, however, a problem since the spectrum 
of a closed (2,2) string contains only {\it one} particle \cite{ov}. A resolution
to that contradiction is possible via the {\it maximal} supersymmetrization of 
SDG, which extends the particle content and delivers the anti-self-dual form 
$\o$, in particular. Simultaneously, it is the (maximal) supersymmetry that 
teaches us how to generalize the $(2,2)$ string theory that we began with, to a 
theory of the M-brane.

First, let me notice that there exists another (dual) description of SDG by 
a {\it five-dimensional} K\"ahler-Lorentz-Chern-Simons action \cite{ket}
$$ I_{\rm KLCS} = -\fracmm{1}{4\p}
\int_Y \o\wedge \Tr\left( \O\wedge R -\fracmm{1}{3}\O\wedge\O\wedge
\O\right)~,\eqno(5)$$
where $Y=M_{2+2}\otimes T$, $M_{2+2}$ is the world-volume of M-brane, $T$ is an 
auxiliary dimension (extra `time' $t$), the dilaton two-form $\o$ is valued on 
the world-volume, $\O$ is the Lorentz Lie algebra-valued one-form on $Y$, and $R$ 
is the associated curvature two-form, $R=d\O+\O\wedge \O$. It is assumed that
the gravitational `spin-connection' $\O$ is related to the K\"ahler metric (2) in
the standard way via a `vierbein', and the trace goes over the Lorentz group.  
The action (5) is formally analoguous to the standard K\"ahler-Chern-Simons 
action for the self-dual Yang-Mills \cite{nsch}. It is straightforward to 
check (e.g., as in ref.~\cite{nsch}) that the equations of motion following from 
the action (5) are actually four-dimensional,~\footnote{ The boundary conditions 
on $\pa Y$ are supposed to be chosen in such a way that no boundary \newline 
${~~~~~}$ terms appear in the equations of motion.}
$$ \o\wedge R=0~,\qquad \fracmm{\pa \O}{\pa t}=0~,\eqno(6)$$
and they can also be obtained from the Donaldson-Nair-Schiff-type action
\cite{don,nsch}
$$ S_{\rm DNS}[J;\o] = -\fracmm{1}{4\p}\int_{M_{2+2}}\,\o\wedge\Tr(J^{-1}\pa J
\wedge J^{-1}\bar{\pa}J) + \fracmm{i}{12\p}\int_{M_{2+2}\otimes T}\,
\o\wedge\tr(J^{-1}dJ)^3 \eqno(7)$$
in terms of the Yang scalar $J$ substituting the connection $\O\,$.

The equations (6) formally describe two world-volume gravities, one being 
self-dual and another being anti-self-dual, so that each serves as a Lagrange 
`multiplier' for another. It is necessary for any covariant off-shell description
of self-duality. In the case of an M-brane, we actually have to deal with the two
spaces: the world-volume which is four-dimensional and the target space whose 
dimension is supposed to be twelve-dimensional. Therefore, by insisting on 
self-duality in the target space too, one can make use of the `second' self-dual 
gravity, but in twelve dimensions. Though the $2+10$ dimensional Lorentz group
$SO(2,10)$ is simple, it is its (minimal) supersymmetric extension that is not
simple, being isomorphic to $OSp(32|1)\otimes OSp(32|1)'$, which allows us to
choose a self-dual factor $OSp(32|1)$ to represent self-dual Lorentz 
supersymmetry in $2+10$ dimensions.
 
When starting from the Pleba\'nski action (1), its {\it world-volume} 
supersymmetrization can be easily performed in the $N$-extended self-dual
superspace \cite{sie,kgn}. Because of the isomorphisms 
$SU(1,1)\cong SL(2,{\bf R})$ and $SO(2,2)\cong SL(2,{\bf R})\otimes 
SL(2,{\bf R})'$, it is natural to represent the $2+2$ `space-time' coordinates
as $z^{\a,\a'}$, where $\a=(+,-)$ and $\a'=(+',-')$ refer to $SL(2)$ and $SL(2)'$,
respectively. The $N$-extended supersymmetrization of self-duality amounts to
extending the $SL(2)$ factor to $OSp(N|2)$, while keeping the $SL(2)'$ one to
be intact. One has $\d^{AB}=(\d^{ab},C^{\a\b})$, where $\d^{ab}$ is the $SO(N)$
metric and $C^{\a\b}$ is the (part of) charge conjugation matrix, $A=(a,\a)$.
In superspace $Z=(z^{\a,\a'},\q^{A'})$, the $N$-extended (gauged) {\it self-dual
supergravity} (SDSG) is defined by the following constraints on the spinorial 
covariant derivatives $\de_{A\a'}=E\du{A\a'}{M\m'}\pa_{M\m'}
+\ha \O_{A\a'BC}M^{CB}$ ~\cite{sie}:
$$ \{ \de^{a\a},\de^{b\b}\}=C^{\a\b}M^{ab}+\d^{ab}M^{\a\b}~,\eqno(8a)$$
$$\{ \de^{a\a},\de_{b\b'}\}=\d^{a}_{b}C^{\a\b}\de_{\b\b'}~,\quad
\[ \de^{a\a},\de_{\b\b'}\]=\d^{\a}_{\b}\d^{ab}\de_{b\b'}~,\eqno(8b)$$
where $M^{AB}=(M^{ab},M^{\a\b},\de^{a\a})$ are the generators of $OSp(N|2)$.
Eqs.~(8) have the $OSp(N|2)\otimes SL(2)'$ (local$\otimes$global) symmetry,
and they can be `solved' in a light-cone gauge, in terms of a SDSG
{\it pre-potential}  $V_{\DM'\DM'}$, where the subscript means `helicity' $(-2)$.
In the light cone-gauge, the SDSG superspace constraints can be reduced to a 
single equation of motion for the super-prepotential, similarly to the bosonic 
case. The equation itself follows from the $N$-extended super-Pleba\'nski action 
whose form is formally $N$-independent and is given by \cite{sie}
$$S_{\rm SDSG}=\int d^{2+2}x d^N\q\left[ \frac{1}{2} V_{\DM'\DM'}\bo V_{\DM'\DM'}
+\frac{i}{6}V_{\DM'\DM'}(\pa\ud{\a}{+'}\pa_{A+'}V_{\DM'\DM'})
\d^{BA}(\pa_{B+'}\pa_{\a+'}V_{\DM'\DM'})\right]~.\eqno(9)$$
As was noticed by Siegel~\cite{sie}, the action (9) implies the {\it maximal}
supersymmetry~! Indeed, dimensional analysis immediately yields $N=8$, and the
same follows from counting the total $GL(1)'$ charge of the action (9), where
$GL(1)'$ is just the unbroken part of the `Lorentz' factor $SL(2,{\bf R})'$. It
is worth mentioning that the $N=8$ ~SDSG also has the hidden non-compact global 
symmetry which is a contraction of the $E_7$ symmetry of the non-self-dual $N=8$
supergravity. Its discrete subgroup may just be the U-duality group of the 
(compactified) F theory ({\it cf.} refs.~\cite{berlin,htow}).

One can `explain' the internal gauged $SO(8)$ symmetry of the $N=8$ ~SDSG as a 
part of the gauged Lorentz symmetry in a higher-dimensional space time, which is 
usual in supergravity. In our case, 
the higher-dimensional spacetime should be $2+10$ dimensional, since 
$SO(2,2)\otimes SO(8)\subset SO(2,10)$. Unlike the $SO(2,2)$, 
the Lie group $SO(2,10)$ is not a product of two self-dual factors, which is yet 
another reason for introducing supersymmetry. The supersymmetric Lorentz symmetry
$OSp(32|1)\otimes OSp(32|1)'$ in $2+10$ dimensions allows us to define the 
self-duality to be associated with the factor $OSp(32|1)$, which is of course 
related to the well-known fact that Majorana-Weyl spinors and self-dual gauge 
fields are also allowed in $2+10$ dimensions.

Unlike the Pleba\'nski-type action (9), the geometrical KLCS-type action (5) 
seems to be more suitable for describing the super-Lorentz self-duality in twelve 
dimensions since it can be generalized to 
$$ S_{\rm M-brane} =-\fracmm{1}{4\p}\int_Y \o \wedge \O_{\rm sLCS}~,\eqno(10)$$
where $\O_{\rm sLCS}\equiv {\rm sTr}\left(\hat{\O}\wedge d\hat{\O}+\fracmm{2}{3}
\hat{\O}\wedge\hat{\O}\wedge\hat{\O}\right)$ is the super-Lorentz-Chern-Simons 
three-form to be constructed in terms of the $OSp(32|1)$ Lie superalgebra-valued 
one-form $\hat{\O}$. The action (10), or its DNS-form similar to that in eq.~(7),
describes the $2+2$ dimensional M-brane with the target space given by the
$528_{\rm B}+32_{\rm F}$ dimensional supergroup \cite{ket}. The world-volume 
two-form $\o$ represents the anti-self-dual gravity or the $(2,2)$ string 
dilaton, whereas the Lie superalgebra-valued one-form $\hat{\O}(z)$, or the 
associated Yang scalar $\hat{J}(z)$, describe the embedding of the M-brane into 
$OSp(32|1)$.
 
The target space of our M-brane is therefore given by a homogeneous supergroup 
manifold, which is obviously not a flat `spacetime'. In particular, the
supersymmetry part of the $OSp(32|1)$ Lie super-algebra reads \cite{hpu}
$$ \{Q_{\a},Q_{\b}\}=\g^{\m\n}_{\a\b}M_{\m\n}+\g^{\m_1\cdots\m_6}_{\a\b}
Z^+_{\m_1\cdots\m_6}~,\eqno(11)$$
where $Q_{\a}$ is a 32-component Majorana-Weyl spinor, $M_{\m\n}$ are 66 Lorentz
generators, 462 bosonic generators $\{Z^+\}$ represent a self-dual six-form, and 
all the Dirac $\g$-matrices are chirally projected. The algebra (11) is not a 
super-Poincar\'e-type algebra because of the absence of translation generators on
 its right-hand-side. Its spacetime interpretation is nevertheless possible in 
lower dimensions. For example, the 66 twelve-dimensional Lorentz generators 
$M_{\m\n}$ can represent 55 Lorentz generators and 11 translation generators 
after a Wigner-In\"on\"u contraction of the superalgebra down to eleven 
dimensions. 

The arguments presented above support the idea that the natural framework for 
describing F theory may be provided by a quantized theory of $2+2$ dimensional 
M-branes `propagating' in a higher-dimensional target space. The self-duality is 
crucial in all the constructions proposed above. Though the twelve-dimensional 
interpretation of the target space is possible, it should be taken formally since
no super-Poincar\'e-type supersymmetry appears. It is also not clear what may be 
the {\it flat} target space limit of the proposed theory (10), since the target 
space there is essentially curved, while the embedding of the M-brane into that
space is described by the self-dual gauge fields. The action (10) is invariant 
under all the relevant symmetries including superymmetry, and it may serve as a
fundamental action of the M-brane. One may speculate about the possibility of the
UV-finiteness of that action~\cite{yale,ket2}, contrary to the naive expectations
based on its formal non-renormalizability.

To conclude, self-duality and supersymmetry seem to be important in determining 
both the world-volume and the target space dynamics of M-brane. Being combined 
with the world-sheet/target space duality between strings and membranes, it 
allowed us to proceed with constructing the M-brane action. The different
(Dirac-Born-Infeld-type) action for the M-brane was proposed from the studies of 
$(2,1)$ heterotic strings in ref.~\cite{kmnew}. The ten-dimensional superstrings 
may be recovered from the M-brane by wrapping a two-dimensional part of its 
world-volume around a compact two-dimensional part of its target space. For the 
time being, however, there are more questions than answers.
\vglue.2in

\end{document}